\documentclass[sn-nature]{sn-jnl}

\usepackage{graphicx}%
\usepackage{multirow}%
\usepackage{amsmath,amssymb,amsfonts}%
\usepackage{amsthm}%
\usepackage{mathrsfs}%
\usepackage[title]{appendix}%
\usepackage{xcolor}%
\usepackage{textcomp}%
\usepackage{manyfoot}%
\usepackage{booktabs}%
\usepackage{algorithm}%
\usepackage{algorithmicx}%
\usepackage{algpseudocode}%
\usepackage{listings}%
\usepackage{siunitx}

\usepackage{steinmetz}
\usepackage{bm}

\usepackage[T1]{fontenc} 
\usepackage{upgreek}

\theoremstyle{thmstyleone}%
\theoremstyle{thmstyletwo}%

\theoremstyle{thmstylethree}%

\begin{document}

\title{Topological Hall Effect Induced by Chiral Spin Textures at the Ferroelectric/Ferromagnetic Interface}


\author[1]{Jingkuan Xiao}
\author[1]{Yaqing Han}
\author[2]{Jianfeng Guo}
\author[1]{Renjun Du}
\author[1,3,4]{Jiawei Jiang}
\author[5]{Baoshan Cui}
\author[2,6]{Runnong Zhou}
\author[1]{Siqin Wang}
\author[1]{Siqi Jiang}
\author[1]{Fuzhuo Lian}
\author[1]{Di Zhang}
\author[1]{Guodong Ma}
\author[1]{Jiabei Huang}
\author[1]{Zhaochen Qu}
\author[1]{Wanting Xu}
\author[7]{Kenji Watanabe}
\author[8]{Takashi Taniguchi}
\author[1]{Alexander S. Mayorov}
\author[1]{Jinsheng Wen}
\author[1]{Haifeng Ding}
\author[1]{Gong Chen}
\author[9]{Ahmet Avsar}
\author[3]{Hongxin Yang}
\author[2,6]{Lihong Bao}
\author[2,6]{Hong-Jun Gao}
\author[2,6]{Shiyu Zhu}
\author[1,10]{Lei Wang}
\author[1,10]{Geliang Yu}






\affil[1]{National Laboratory of Solid State Microstructures, Collaborative Innovation Center of Advanced Microstructures, School of Physics, Nanjing University, Nanjing, China}

\affil[2]{Beijing National Center for Condensed Matter Physics and Institute of Physics, Chinese Academy of Sciences, Beijing, China}

\affil[3]{Center for Quantum Matter, School of Physics, Zhejiang University, Hangzhou, China}

\affil[4]{Department of Chemistry, School of Science, Tianjin University, Tianjin, China}

\affil[5]{Key Laboratory for Magnetism and Magnetic Materials of Ministry of Education, Lanzhou University, Lanzhou, China}

\affil[6]{School of Physical Sciences, University of Chinese Academy of Sciences, Beijing, China}

\affil[7]{Research Center for Electronic and Optical Materials, National Institute for Materials Science, Tsukuba, Japan}

\affil[8]{Research Center for Materials Nanoarchitectonics, National Institute for Materials Science, Tsukuba, Japan}

\affil[9]{Department of Materials Science and Engineering, National University of Singapore, Singapore}

\affil[10]{Jiangsu Physical Science Research Center, Nanjing University, Nanjing, China}


\affil{$^{\dagger}$These authors contributed equally to this work}
\affil{$^{\ast}$Corresponding authors. E-mails: mayorov@nju.edu.cn; syzhu@iphy.ac.cn; leiwang@nju.edu.cn; yugeliang@nju.edu.cn}






\abstract{Chiral spin textures, largely driven by the Dzyaloshinskii-Moriya interaction, offer significant potential for next-generation computing technologies due to their chirality and topological stability.
Ferroelectric/ferromagnetic van der Waals heterostructures are particularly appealing because they can combine interfacial inversion-symmetry breaking and spin-orbit coupling to promote interfacial Dzyaloshinskii-Moriya interaction, while switchable ferroelectric polarization provides a nonvolatile tuning knob.
This study investigates interfacial chiral spin textures in few-layer Fe$_3$GeTe$_2$/$\alpha$-In$_2$Se$_3$ heterostructures.
Two groups of topological Hall signals are identified just below and above the coercive field, and thickness-dependent transport reveals a notable reduction in critical temperature with increasing Fe$_3$GeTe$_2$ layer thickness.
Low-temperature magnetic force microscopy images reveal two types of magnetic bubbles with opposite magnetic contrasts near the coercive field, each associated with distinct topological Hall signals.
Together with atomistic spin-dynamics simulations and first-principles calculations, these results support the formation of interfacial DMI-stabilized chiral spin textures.
Switching the ferroelectric polarization of the $\alpha$-In$_2$Se$_3$ layer further enables nonvolatile modulation of both anomalous and topological Hall effects. 
The resulting ferroelectric and magnetic bistabilities generate four distinguishable Hall resistance states programmable by electric and magnetic fields.  
These findings highlight the potential of van der Waals interfaces for advanced device applications.}







\maketitle







\section{Introduction}

Two-dimensional (2D) van der Waals (vdW) magnets provide an atomically thin platform for manipulating magnetic order and spin-dependent transport~\cite{Park2026RMP, Antonija20252DM}. 
Their dangling-bond-free surfaces, atomically sharp interfaces, and mechanical stackability of vdW materials enable the assembly of designed magnetic heterostructures with highly tunable interfacial functionalities~\cite{Gibertini2019, Cheng2019S, burch2018}. 
Interfacial engineering in such heterostructures offers versatile routes for tailoring magnetic interactions through interfacial electronic reconstruction, exchange coupling, and structural reconstruction~\cite{Mellado2025, Pan2024NN, Jia2025AN}. 
Consequently, vdW heterointerfaces provide an effective platform for engineering magnetic properties, enabling the modulation of magnetic anisotropy, coercivity, and magnetic ordering through interfacial coupling~\cite{Rizzo2025, Fujita2024AFM, Ma_2024, Wang2025AN}.
Furthermore, recent studies have established twist-induced moir\'{e} reconstruction as an additional structural degree of freedom for engineering spatially modulated magnetic interactions and emergent spin textures in CrI$_3$- and Fe$_3$GeTe$_2$ (FGT)-based systems~\cite{Cheng2023NE, Yang2024NC, Ye2025PRL, Kim2026NC}. 
Beyond structural engineering, ferroelectricity introduces an electrical degree of freedom for actively controlling interfacial magnetism in vdW heterostructures.
Switchable polarization can reversibly modulate interfacial charge redistribution and symmetry, thereby tuning magnetic interactions and spin-dependent transport~\cite{Eom2023NatCommun, Liang2026NC, Xie2026NC}.
This provides a route toward electrically reconfigurable chiral magnetism and associated transport phenomena.

Among the interfacial interactions accessible in vdW heterostructures, the Dzyaloshinskii-Moriya interaction (DMI), which arises from spin-orbit coupling in systems with broken inversion symmetry, is particularly important for forming and stabilizing chiral spin textures, including skyrmions, chiral magnetic bubbles, and chiral domain walls~\cite{DZYALOSHINSKY1958241, Moriya1960, Manchon2015}. 
When conduction electrons traverse such chiral spin textures, they acquire a real-space Berry phase, which can be described as an emergent magnetic field and gives rise to the topological Hall effect (THE)~\cite{Nagaosa2013NatNano,fert2017}.
Considerable efforts have therefore been devoted to engineering interfacial DMI and exploring Hall signatures associated with chiral spin textures in FGT-based vdW systems, including WTe$_2$/FGT~\cite{Wu2020NC}, FGT/Cr$_2$Ge$_2$Te$_6$~\cite{Wu2022fd}, and twisted FGT structures~\cite{Kim2026NC}.
Despite this progress, a comprehensive understanding of the microscopic mechanisms responsible for their formation and stabilization, together with the realization of active, reversible, and nonvolatile electrical control of the THE, remains a major challenge in 2D magnetic systems~\cite{Zhong2025IOP, WANG2022100971}. 
Furthermore, direct real-space observation of chiral spin textures in air-sensitive few-layer vdW heterostructure devices remains experimentally demanding. 
Addressing these challenges is essential for harnessing electrically reconfigurable vdW magnetic heterostructures toward next-generation low-power spintronic and multistate memory technologies~\cite{Zhang2024NS, Wu2020NC}.

Here, we report a ferroelectric/ferromagnetic (FE/FM) vdW heterostructure based on the itinerant ferromagnet FGT and room-temperature ferroelectric $\alpha$-In$_2$Se$_3$ (IS), which provides a tunable platform for exploring interfacial chiral spin textures. 
We observe two groups of topological Hall signals emerging on opposite sides of the coercive field, indicating a nontrivial evolution of spin textures during magnetization reversal. 
By combining thickness-dependent magnetotransport measurements, magnetic force microscopy (MFM) imaging, atomistic spin-dynamics simulations, and first-principles calculations, we attribute these two topological Hall components to interfacial chiral spin textures stabilized by a finite DMI.
Furthermore, we demonstrate nonvolatile tuning of both the anomalous Hall effect (AHE) and the THE through ferroelectric polarization switching of the IS layer.
By leveraging the coupled ferroelectric and ferromagnetic bistabilities, we realize a nonvolatile four-state memory functionality, highlighting the potential of FE/FM vdW interfaces for tunable topological spintronics and multibit devices.

\section{Results and Discussion}

\begin{figure*}[htb]
	\includegraphics[width=\textwidth]{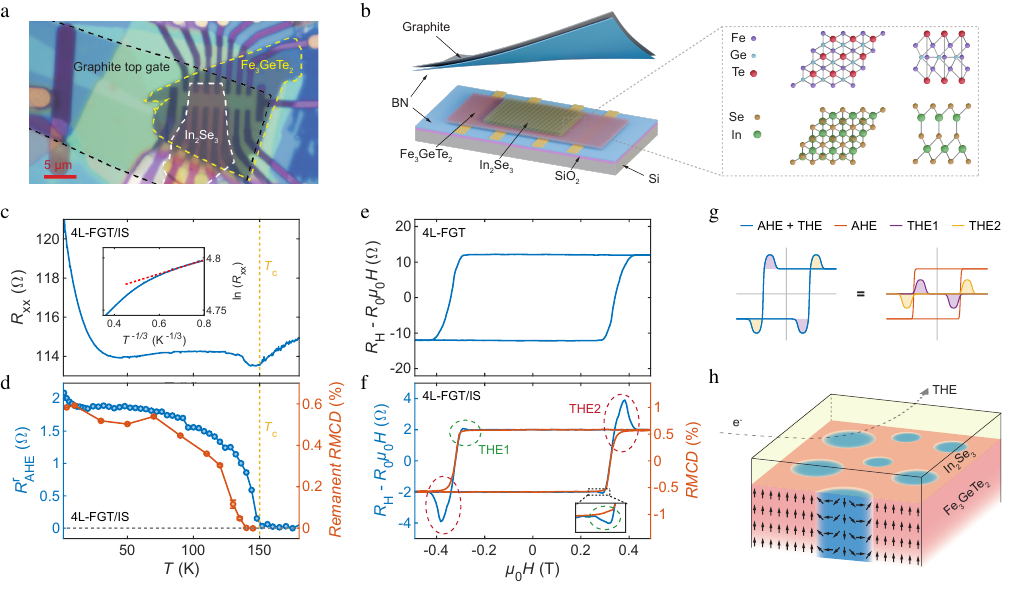}
	\caption{Sample structure and transport properties.
		a) Optical micrograph of the fabricated device. The scale bar corresponds to $5\,\upmu\mathrm{m}$.
		b) Schematic illustration of the FGT/IS vdW heterostructure device. The zoom-in displays the top and side views of monolayer FGT and IS.
		c) Temperature-dependent longitudinal resistance of the FGT/IS heterostructure. Inset: Fit to the variable-range hopping model, $R_\mathrm{xx} \propto \exp[(T_0/T)^{1/3}]$.
		d) Temperature dependence of the remanent anomalous Hall resistance $R_\mathrm{AHE}^\mathrm{r}$ (blue solid line) and remanent RMCD signal (orange solid line) in the FGT/IS heterostructure. The yellow dashed line indicates the ferromagnetic transition temperature $T_\mathrm{c}$, determined from where $R_\mathrm{AHE}^\mathrm{r}$ drops to zero. The error bars correspond to the standard deviation of the noise in the RMCD signal.
		e) Magnetic field dependence of anomalous Hall resistance measured in the pristine FGT region.
		f) Magnetic field dependence of the Hall resistance after subtracting the ordinary Hall background ($R_\mathrm{H} - R_\mathrm{0}\mu_\mathrm{0} H$) at $T = 4\,\mathrm{K}$ in the FGT/IS heterostructure. The AHE contribution (orange solid line) is confirmed via the RMCD signal.
		g) Schematic illustration showing how non-monotonic Hall signals can be interpreted as a superposition of an AHE and two THE.
		h) Schematic of the magnetic structure in the FGT/IS vdW heterostructure.
		\label{Fig_1}}
\end{figure*}

We fabricated a dual-gated device based on an FGT/IS vdW heterostructure, as shown in \textbf{Figure~\ref{Fig_1}}a, b.
FGT and IS both adopt hexagonal crystal structures with similar in-plane lattice constants of 4.029 and 4.063$\,\si{\angstrom}$, respectively~\cite{Michael2018, May2016}, corresponding to a small lattice mismatch of 0.8\,$\%$.
We first measured the zero-field temperature dependence of the longitudinal resistance $R_\mathrm{xx}$ for the 4L-FGT/IS device.
As shown in Figure~\ref{Fig_1}c, $R_\mathrm{xx}$ exhibits metallic behavior at high temperatures and a distinct kink near 150\,K, indicating the ferromagnetic phase transition.
Furthermore, we characterized the temperature dependence of the remanent anomalous Hall resistance $R_\mathrm{AHE}^\mathrm{r}$ and the reflectance magneto-circular dichroism (RMCD) signal (Figure~\ref{Fig_1}d) in the same region.
The remanent anomalous Hall resistance $R_\mathrm{AHE}^\mathrm{r}$, which is proportional to the spontaneous magnetization, vanishes at the Curie temperature $T_\mathrm{c}\sim$150\,K, while the RMCD signal yields a slightly lower $T_\mathrm{c}$ of $\sim$140\,K.
These results are quantitatively consistent with prior reports on the 4L-FGT~\cite{2018Deng563, Fei2018}.
At lower temperatures ($T < 40\,\mathrm{K}$), $R_\mathrm{xx}$ exhibits an upturn characteristic of insulating behavior (inset of Figure~\ref{Fig_1}c), which can be well-described by a 2D variable-range hopping model ($R_\mathrm{xx} \propto \exp[(T_0/T)^{1/3}]$), indicating carrier localization at low temperatures~\cite{Liu2017, Roemer2020}.

To investigate the magnetic properties of the FGT/IS heterostructure, we measured the Hall resistance $R_\mathrm{H}$ under a perpendicular magnetic field.
To eliminate any longitudinal resistance $R_\mathrm{xx}$ contributions arising from geometric misalignment of the pre-patterned contacts, the measured Hall resistance was antisymmetrized with respect to the magnetic field (see Methods).
In pristine FGT, the total Hall resistance can be expressed as $R_\mathrm{H} = R_\mathrm{0}\mu_\mathrm{0} H + R_\mathrm{AHE}$, where $R_\mathrm{0}$ is the ordinary Hall coefficient (extracted from the high-field linear regime) and $R_\mathrm{AHE}$ is the anomalous Hall contribution.
As depicted in Figures~\ref{Fig_1}e and f, after subtracting the linear ordinary Hall background, the isolated AHE signal manifests as a square-shaped hysteresis loop with a coercive field of $\mu_0H_\mathrm{c} = 0.34\,\mathrm{T}$ at $4\,\mathrm{K}$, indicating robust long-range ferromagnetic order with strong perpendicular magnetic anisotropy. This characteristic behavior is consistently observed across both pristine FGT and FGT/IS devices of varying thicknesses (Figure~S4, Supporting Information).

\begin{figure*}[htbp]
	\includegraphics[width=\textwidth]{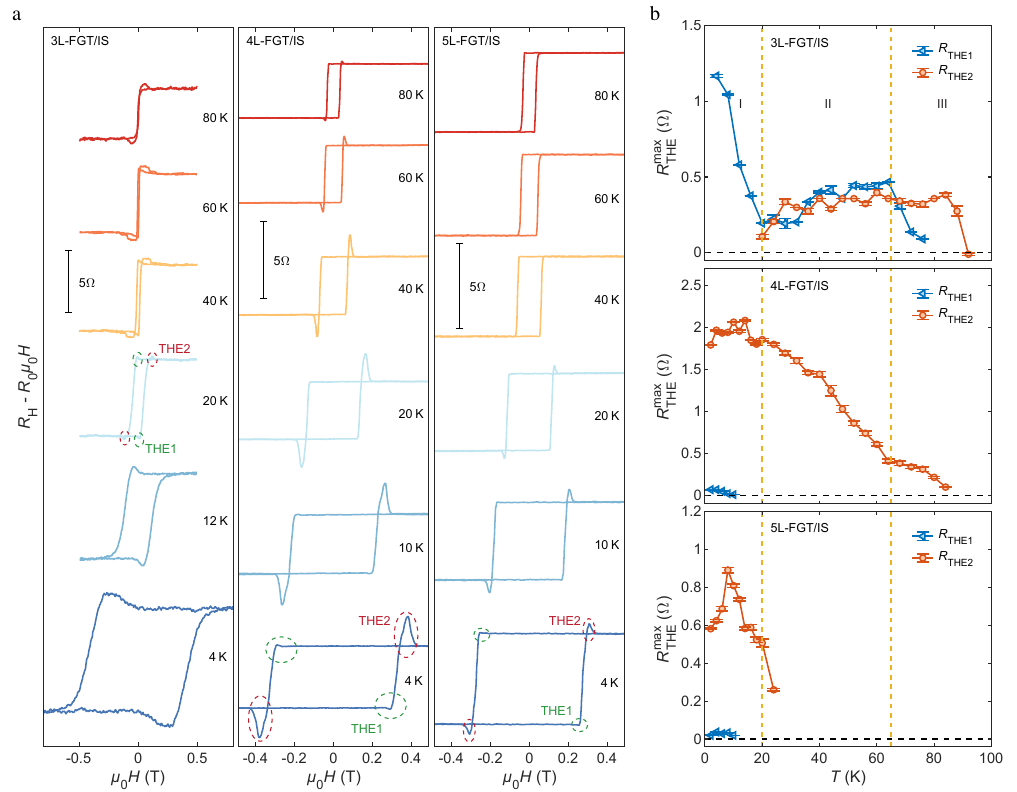}
	\caption{Temperature dependence of two groups of THE.
		a) Magnetic field dependence of the anomalous and topological Hall resistance at various temperatures in the FGT/IS heterostructures. All curves are antisymmetrized to eliminate contributions from the longitudinal resistance, and the ordinary Hall component is subtracted via linear fitting in the high-field regime. Green and red dashed circles indicate the positions of the topological Hall features, labeled as THE1 and THE2.
		b) Temperature dependence of the maximum topological Hall resistance $R_\mathrm{THE}^\mathrm{max}$. The error bars correspond to the standard deviation of the noise in the respective measurements. \label{Fig_2}}
\end{figure*}

In contrast to pristine FGT, we observe two groups of humps during magnetization reversal (Figure~\ref{Fig_1}f), appearing below and above the coercive field, respectively.
These nontrivial Hall features can be attributed to either an intrinsic THE or the superposition of two AHE signals with reversed signs~\cite{Wu2020NC, Wu2022fd, Kimbell2020PRM, Gerber2018PRB, Wang2020NL}.
After rigorously excluding the latter scenario through additional analyses of parallel conduction channels and magnetic inhomogeneities (Section~3 in the Supporting Information), these analyses strongly support a topological origin of the signals.
To precisely extract the $R_\mathrm{THE}$, we performed RMCD measurements on the same region to determine the AHE background.
By subtracting this AHE background, the residual Hall signal can be decomposed into two distinct THE components with opposite signs, labeled THE1 and THE2 (schematically illustrated in Figure~\ref{Fig_1}g).
Because IS is highly resistive and nonmagnetic, the observed THE is attributed to interfacial chiral spin textures in FGT, such as skyrmions or skyrmion bubbles, which generate an emergent magnetic field and a real-space Berry phase for conduction electrons, as illustrated in Figure~\ref{Fig_1}h.

We further investigated the dependence of the two THE components on the FGT thickness.
As shown in \textbf{Figure~\ref{Fig_2}}a, THE1 and THE2 are distinctly observed in the 3L-, 4L-, and 5L-FGT/IS heterostructures.
For the 3L-FGT/IS device, the topological Hall response can be divided into three temperature regimes (Figure~\ref{Fig_2}b).
In Region I ($T < 20\,\mathrm{K}$), THE1 decreases with increasing temperature, whereas THE2 is absent. In Region II ($20\,\mathrm{K} < T < 65\,\mathrm{K}$), both THE1 and THE2 remain nearly unchanged with increasing temperature. In Region III ($T > 65\,\mathrm{K}$), both signals progressively weaken with increasing temperature.
The critical temperature $T_\mathrm{c}$ for THE1 is approximately $80\,\mathrm{K}$ in the 3L-FGT/IS device, but falls below $20\,\mathrm{K}$ in both the 4L- and 5L-FGT/IS devices.
Similarly, the critical temperatures of THE2 are $92\,\mathrm{K}$, $84\,\mathrm{K}$, and $24\,\mathrm{K}$ for the 3L-, 4L-, and 5L-FGT/IS heterostructures, respectively.

In our FGT/IS heterostructure, the simultaneous breaking of both the (001) mirror plane reflection ($\widehat{M}_\mathrm{\bot}$) and the combined time-reversal and two-fold rotational symmetry ($\widehat{T}\widehat{C}_\mathrm{2\parallel}$) generates the interfacial DMI~\cite{Huang2022nl, Huang2019prb}.
The estimated twist angles are $\geq 9^\circ$, corresponding to moiré periods below 3\,nm, making long-wavelength moiré reconstruction unlikely to dominate the observed THE (see Section~1.2 of the Supporting Information).
The rapid suppression of the THE with increasing thickness indicates an intrinsically interfacial origin, highlighting a competition between the interfacial DMI and bulk ferromagnetism~\cite{Chen2013, Zhao2018NanoRes, Lv2024AdvFunctMater}.

First-principles calculations for 1L--5L FGT/IS heterostructures reveal a thickness-dependent evolution of the competing magnetic interactions (Figure~S9 in the Supporting Information).  
With increasing FGT thickness, $|D_N|$ decreases continuously, whereas $J_N$ increases markedly between 3L and 4L, shifting the magnetic-energy balance toward bulk ferromagnetism.  
This thickness-dependent evolution is consistent with the markedly reduced THE1 disappearance temperature, from $\sim80\,\mathrm{K}$ in 3L to below $20\,\mathrm{K}$ in 4L.  
To characterize the relative balance among DMI, exchange, and magnetocrystalline anisotropy, we define  
\begin{equation}  
\kappa_N=\frac{\pi|D_N|}{4\sqrt{J_NK_{\mathrm{MCAE},N}}}.  
\end{equation}  
The decrease of $\kappa_N$ with thickness indicates a progressively weaker relative contribution of the interfacial DMI and is consistent with a crossover toward a regime increasingly governed by bulk-like ferromagnetic interactions.

The THE2 amplitude exhibits a non-monotonic thickness dependence, reaching its maximum in the 4L device.
This enhancement cannot be attributed solely to $R_0$, because $|R_{\mathrm{THE2}}/R_0|$ at $20\,\mathrm{K}$ is larger in the 4L device ($\sim$6.81\,T) than in the 3L ($\sim$2.26\,T) and 5L ($\sim$4.12\,T) devices.
Instead, it likely reflects the interplay between the thickness-dependent energetics of chiral textures and the nonequilibrium evolution of domains during magnetization reversal. 
Previous real-space studies have shown that texture formation in FGT depends on temperature, field history, and the reversal pathway~\cite{Birch2022}. Such kinetic effects may also account for the absence of THE2 below $20\,\mathrm{K}$ in the 3L device, where stronger low-temperature pinning and restricted domain-wall mobility may reduce the population or lifetime of THE2-associated textures below the detection threshold.This interpretation is also consistent with the temperature evolution of THE2 in the 4L and 5L devices. Moderate warming may facilitate texture formation, whereas stronger thermal fluctuations suppress the signal at higher temperatures.

The pronounced THE2 response in the 4L device may therefore arise from a favorable balance between the remaining interfacial DMI, bulk ferromagnetic interactions, and texture-formation dynamics.
At 5L, the stronger bulk-ferromagnetic contribution further weakens the relative role of the interfacial DMI and narrows the chiral-texture stability window, consistent with the lower THE2 disappearance temperature of $\sim$24\,K.
Specifically, although $J_N$ remains strong at $\sim$18.6\,meV/f.u., $|D_N|$ decreases by approximately $66\%$ from 4L to 5L for the $P^{\uparrow}$ state, resulting in a substantial reduction of $\kappa_N$.
The 5L heterostructure thus approaches a critical-thickness crossover regime in which the interfacial DMI becomes progressively less effective relative to exchange and anisotropy.

\begin{figure*}[ht]
  \includegraphics[width=1\textwidth]{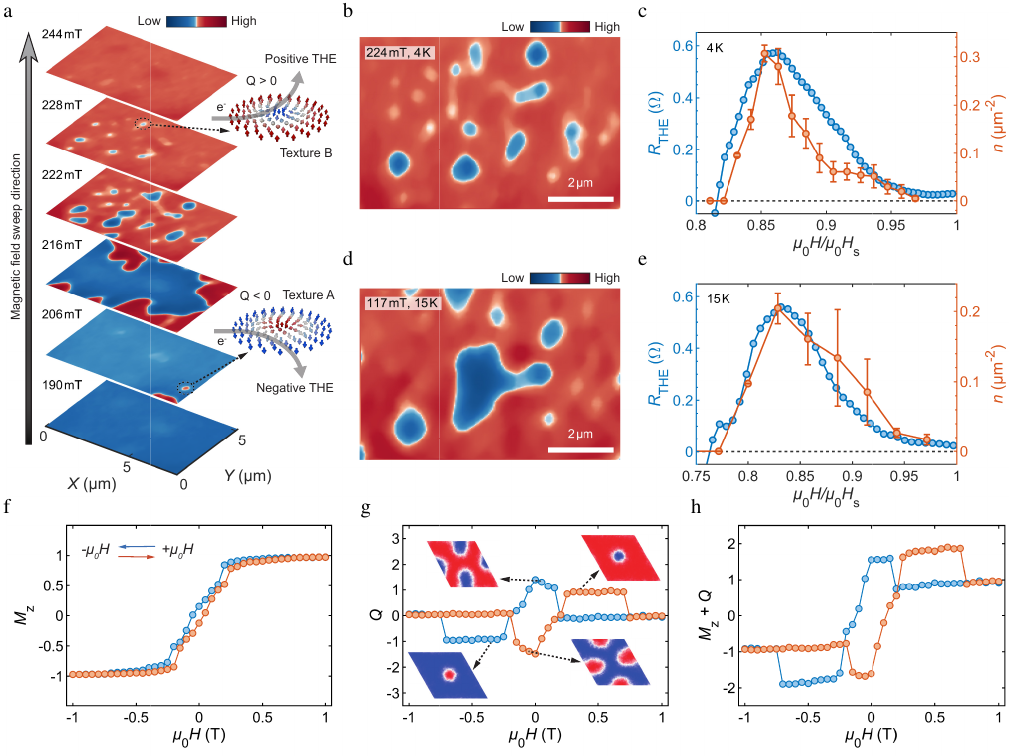}
	\caption{Correlation between magnetic bubbles and the THE.
		a) Field-dependent MFM images of the 5L-FGT/IS heterostructure at $4\,\mathrm{K}$, acquired after applying a $-300\,\mathrm{mT}$ reset field.
    The images reveal two distinct types of bubbles, distinguished by their MFM contrast (red and blue, marked by black dashed circles), which correspond to opposite out-of-plane core magnetizations.
    Notably, bubbles with red contrast emerge at lower fields, while those with blue contrast dominate at higher fields.
    Schematics illustrating the THE sign reversal associated with topological charges $Q > 0$ and $Q < 0$.
    b and d) MFM images of the 5L-FGT/IS heterostructure at 4\,K (b) and 15\,K (d), taken after applying a reset field of -300\,mT.
		c and e) Comparison of the topological Hall resistance (blue circles, left axis) and the corresponding magnetic bubble density (orange circles, right axis) as a function of magnetic field at 4\,K (c) and 15\,K (e).
    The data show a clear correlation between the emergence of bubbles and the topological Hall signal.
    The error bars on the bubble density arise from the image analysis procedure. $\mu_\mathrm{0}H_\mathrm{s}$ is the magnetic field at which the THE and magnetic bubbles vanish.
   f) Calculated out-of-plane magnetization $M_\mathrm{z}$ as a function of the magnetic field. The blue (orange) line represents the magnetic-field-sweeping direction from positive (negative) to negative (positive). 
   g)  Simulated evolution of the topological charge $Q$ along the hysteresis loop depicted in (f). The insets detail the simulated chiral spin textures.
   h) Concurrent evolution of the topological charge $Q$ and $M_\mathrm{z}$ during the simulated magnetic reversal process.
		\label{Fig_3}}
\end{figure*}

After identifying an interfacial origin linked to transport, we now seek direct real-space evidence of the underlying spin textures.
We utilize low-temperature MFM to visualize these real-space spin textures on the vdW interface and clarify the origin of the two groups of THE signals.
After saturating the sample with a negative field of $-300$\,mT at 4\,K, we swept the field in the positive direction (\textbf{Figure~\ref{Fig_3}}a).
The MFM images reveal a remarkable field-driven evolution of magnetic bubbles.
At lower fields (e.g., $\sim$206\,mT), bubbles with an upward-pointing core (red contrast) emerge from a downward-magnetized background (blue contrast).
As the field increases and passes the coercive point (e.g., $\sim$222\,mT), a high density of bubbles with a downward-pointing core (blue contrast) within an upward-magnetized background (red contrast) emerges. The density of these blue-contrast bubbles progressively decreases as the field increases further, vanishing beyond ~240\,mT.

To quantify the bubble dimensions, we converted the original MFM images (Figure~\ref{Fig_3}b) to binary images and used image analysis software to count individual bubbles (Figure~S13, Supporting Information).
The extracted bubble sizes range from 0 to 0.9$\,\upmu\mathrm{m}^2$, with an average area of approximately 0.3$\,\upmu\mathrm{m}^2$. 
We then extracted the field-dependent density of blue-contrast bubbles ($n = counts/S$, where $counts$ is the number of blue bubbles and $S$ is the scanned area) and plotted it alongside the THE signal (Figure~\ref{Fig_3}c, right axis).
A striking correlation emerges: the blue bubble population closely tracks the THE2 signal.
Even at $T = 15\,\mathrm{K}$ (Figure~\ref{Fig_3}d), where thermal effects reduce the overall bubble density, the strong correlation between THE2 and the blue-contrast bubble population persists (Figure~\ref{Fig_3}e).
This one-to-one correspondence provides direct, real-space evidence that the evolution of these magnetic bubbles is responsible for the observed topological Hall signals.

Unlike dual-THE phenomena reported in multi-layer or multi-channel systems~\cite{Wu2022fd, Wang2021NL}, our FGT/IS heterostructure comprises only a single magnetic constituent (FGT).
While recent work on synthetic antiferromagnets has shown that dual-polarity skyrmions can be stabilized by competing interlayer fields~\cite{Fallon2026}, our single-layer system employs a distinct mechanism. 
To elucidate this, atomistic spin-dynamics simulations were performed using first-principles parameters (see Methods). 
As shown in Figure~\ref{Fig_3}f, the simulated out-of-plane magnetization $M_\mathrm{z}$ exhibits the characteristic hysteretic behavior of monolayer FGT~\cite{2018Deng563}.
More importantly, the field-dependent topological charge in Figure~\ref{Fig_3}g exhibits two extrema of opposite sign on either side of the coercive field. 
During the negative-to-positive field sweep, chiral textures with one core--background polarity nucleate below the coercive field and produce a negative $Q$. 
After the background magnetization reverses, chiral textures with reversed core and background magnetizations emerge above the coercive field, producing a positive $Q$. 
The sequence is reversed during the positive-to-negative field sweep.

Based on the combined transport, MFM, and atomistic spin-dynamics results, we propose the following physical picture for the dual THE.
Below the coercive field ($\mu_0H<\mu_0H_\mathrm{c}$), red-contrast bubbles with one core--background polarity emerge (Texture A) and are associated with the negative THE1 signal.
The weak THE1 response in the 5L-FGT/IS device is consistent with the low abundance of red-contrast bubbles within the MFM scan area.  
Above $\mu_0H_\mathrm{c}$, these bubbles give way to blue-contrast bubbles with the reversed core--background polarity (Texture B), whose field-dependent population closely tracks the positive THE2 signal.
The sign reversal of $Q$ can be understood from $Q\propto pv$, where $p$ and $v$ denote the core polarity and vorticity, respectively. 
The interfacial Dzyaloshinskii--Moriya interaction selects a preferred chirality, while the simulations indicate that the vorticity is preserved during magnetic reversal.
Reversal of the core--background polarity across the coercive field therefore reverses the sign of $Q$.
Thus, the dual THE is attributed to a field-driven reversal of the core--background polarity and the associated topological charge within the same FGT magnetic subsystem.

This interpretation is further supported by Figure~\ref{Fig_3}h, where the concurrent evolution of $M_\mathrm{z}$ and $Q$ captures the key features of the experimentally observed Hall response.
Although the simulated textures below $\mu_0H_\mathrm{c}$ differ morphologically from the experimentally observed bubble-like states, this difference may reflect spatially nonuniform interfacial conditions that are absent from the idealized simulations, including local strain, structural disorder, and defects.
Interfacial strain may play an important role in the present heterostructure, given the reported strain-mediated modulation of magnetism in FGT/IS heterostructures~\cite{Eom2023NatCommun, Fujita2024AFM}.
Moreover, the monolayer FGT simulated here is expected to be more susceptible to interfacial strain than the experimentally measured 5L FGT, which may further contribute to the difference in texture morphology.
By locally modifying the magnetic anisotropy, exchange interactions, and Dzyaloshinskii--Moriya interaction, spatially varying strain could alter the balance between different chiral textures and favor bubble-like configurations~\cite{Tanaka2020PRM}.
Importantly, the characteristic size of the chiral spin textures estimated from the THE is comparable to that obtained from atomistic spin-dynamics simulations (see Section~6 of the Supporting Information), supporting the consistency between the simulated and experimentally inferred textures.

\begin{figure*}[htbp]
	\includegraphics[width=\textwidth]{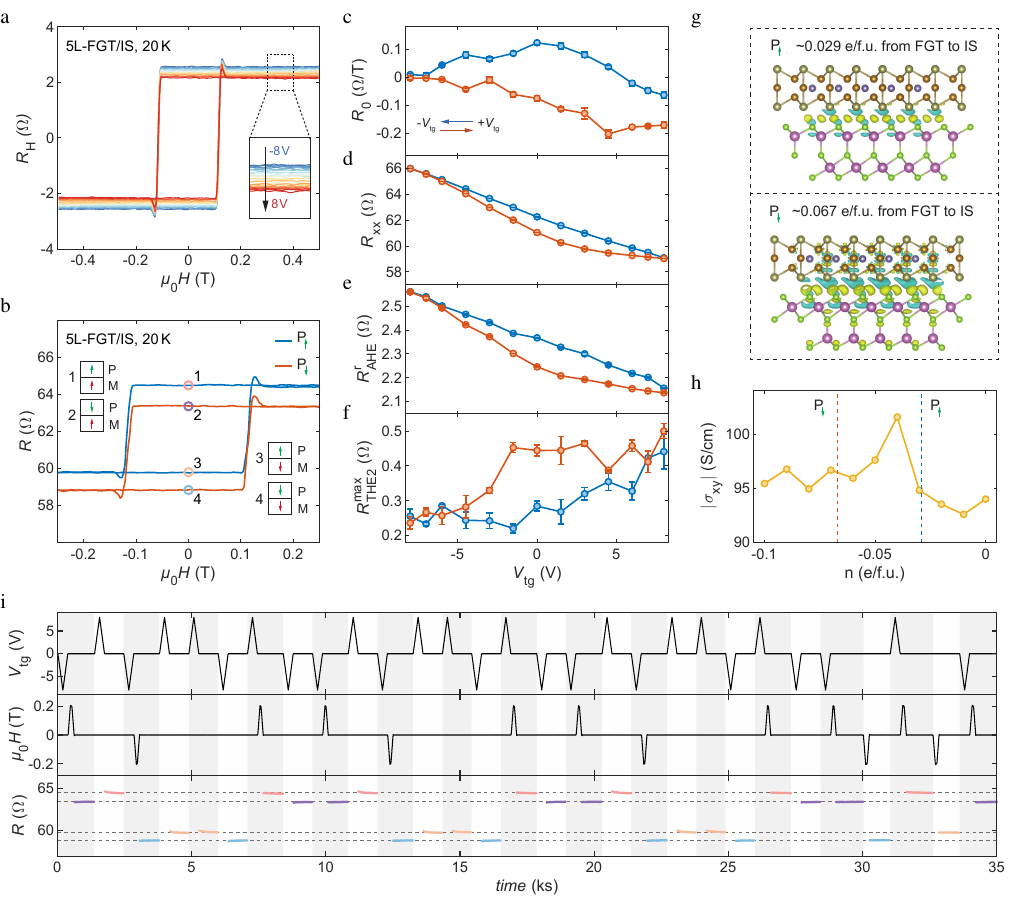}
	\caption{Ferroelectric control of anomalous and topological Hall effects.
		a) Magnetic field dependence of the Hall resistance under various top-gate voltages in 5L-FGT/IS heterostructure at $T = 20\,\mathrm{K}$.
		b) Magnetic field dependence of the resistance under positive poling (blue solid line) and negative poling (orange solid line). Positive (negative) poling corresponds to sweeping the top gate from +8\,V ($-8$\,V) to 0\,V. The circles around the zero magnetic field highlight four distinct resistance states. Green (red) arrows mark the direction of electric (magnetic) field poling.
		c-f) Top-gate voltage dependence of the ordinary Hall coefficient $R_0$ (c), longitudinal resistance $R_\mathrm{xx}$ (d), remanent anomalous Hall resistance $R_\mathrm{AHE}^\mathrm{r}$ (e) and maximum topological Hall resistance $R_\mathrm{THE}^\mathrm{max}$ (f). The error bars correspond to the standard deviation of the noise in the respective measurements. The blue (orange) circles and solid lines represent the top-gate voltage sweep from +8\,V to $-8$\,V ($-8$\,V to +8\,V).
		g) First-principles calculations of the atomic structures and corresponding differential charge densities for the FGT/IS heterostructure under opposite FE polarizations. The cyan and yellow colors at the interfaces represent charge accumulation and depletion, respectively. The electron doping levels in the 2D FGT induced by up ($\textrm{P}_{\uparrow}$) and down ($\textrm{P}_{\downarrow}$) polarizations are $\sim$0.029\,e/f.u. and $\sim$0.067\,e/f.u., respectively.
    h) Calculated Hall conductivity $\sigma_{xy}$ of the FGT/IS heterostructure as a function of electron doping density. The red and blue dashed lines correspond to the calculated doping levels induced by down ($\textrm{P}_{\downarrow}$) and up ($\textrm{P}_{\uparrow}$) FE polarizations, respectively.
		i) Time-dependent switching behavior among the four resistance states. The four states are indicated by dashed lines in pink (1), purple (2), orange (3), and blue (4). \label{Fig_4}}
\end{figure*}

Having established the origin of these chiral spin textures, we now turn to the ferroelectric manipulation of the AHE and THE in the FGT/IS heterostructure.
While the high carrier density of itinerant ferromagnetic FGT typically restricts electrostatic modulation to ionic liquid gating~\cite{2018Deng563}, our heterostructures achieve an effective gate-dependent response directly through the proximity of the ferroelectric IS layer.
As shown in \textbf{Figure~\ref{Fig_4}}a, the Hall resistance $R_\mathrm{H}$ in the 5L-FGT/IS device gradually decreases as the top-gate voltage $V_\mathrm{tg}$ is swept from $-8$\,V to $+8$\,V at $V_\mathrm{bg} = 0$. 
Notably, owing to the ferroelectric nature of the IS layer, the FGT/IS heterostructure exhibits distinct resistance loops under positive poling ($\textrm{P}_{\uparrow}$, sweeping from +8\,V to 0\,V) and negative poling ($\textrm{P}_{\downarrow}$, sweeping from $-8$\,V to 0\,V), as illustrated in Figure~\ref{Fig_4}b.
This remanent response indicates that the Hall transport can be controlled by the FE polarization state even after the external gate bias is removed.

To further characterize this FE-controlled transport response, we extracted the ordinary Hall coefficient $R_0$, longitudinal resistance $R_\mathrm{xx}$, remanent anomalous Hall resistance $R_\mathrm{AHE}^\mathrm{r}$, and maximum topological Hall resistance $R_\mathrm{THE}^\mathrm{max}$ as functions of $V_\mathrm{tg}$ (Figure~\ref{Fig_4}c-f).
When $V_\mathrm{tg}$ is swept forward and backward between $-8$\,V and $+8$\,V, all these quantities exhibit clear and fully reversible hysteresis loops.
In contrast, the hysteresis is nearly completely suppressed when sweeping the back-gate voltage $V_\mathrm{bg}$, as expected because the IS layer is located on top of FGT and the metallic FGT layer effectively screens the back-gate electric field (Figure~S14 in the Supporting Information).
Similar gate-tunable behavior is also observed in the 3L- and 4L-FGT/IS heterostructures (Figure~S15 in the Supporting Information). In contrast, it is absent in pristine FGT devices under identical gate voltages (Figure~S16 in the Supporting Information), confirming that the ferroelectric IS layer is essential for achieving the observed nonvolatile Hall modulation.

As depicted in Figure~\ref{Fig_4}c, $R_0$ changes systematically during the $V_\mathrm{tg}$ sweep, indicating that the FE polarization modifies the effective ordinary Hall response of the interfacial FGT layer. 
The maximum topological Hall resistance $R_\mathrm{THE}^\mathrm{max}$ increases with $|R_0|$, consistent with the theoretical scaling relation $R_\mathrm{THE} \propto |R_0|$~\cite{Wang2018}.
In contrast, the remanent anomalous Hall resistance $R_\mathrm{AHE}^\mathrm{r}$ exhibits a distinct gate dependence that cannot be attributed solely to the change in $R_0$. 
To identify the dominant contribution in the present devices, we further examined the magnetic properties under gating. 
As shown in Figures~S17 and S18 in the Supporting Information, the Curie temperature $T_\mathrm{c}$ and the overall magnetization $M$ probed by RMCD measurements remain nearly unchanged upon FE switching. 
Thus, according to $R_\mathrm{AHE}=R_S M$, the observed modulation of $R_\mathrm{AHE}^\mathrm{r}$ is mainly associated with changes in $R_S$, rather than with a change in magnetization.
Previous studies have shown that appreciable strain can measurably modify the ferromagnetic properties of FGT/IS heterostructures~\cite{Fujita2024AFM, Eom2023NatCommun}. In our devices, however, neither $T_\mathrm{c}$ nor the RMCD-probed magnetization exhibits a resolvable change upon FE switching, suggesting that strain-mediated modulation is unlikely to be the dominant origin of the observed FE-dependent AHE response.

To uncover the microscopic origins of this behavior, we performed first-principles calculations on the FGT/IS heterostructure.
As depicted in Figure~\ref{Fig_4}g, the interfacial charge transfer is highly sensitive to the FE polarization, arising from a pronounced difference in wavefunction overlap across the interface between the $\textrm{P}_{\uparrow}$ and $\textrm{P}_{\downarrow}$ states, which results in distinct electron doping levels of $\sim$0.029\,e/f.u. and $\sim$0.067\,e/f.u., respectively.
Such polarization-dependent interfacial electronic reconstruction demonstrates that the FE polarization of IS provides an effective nonvolatile control parameter for tuning the electronic environment of the magnetic FGT layer~\cite{Liang2026NC}.
The resulting interfacial charge redistribution reshapes the local electronic structure of the interfacial FGT layer, thereby altering the Berry curvature distribution that governs $R_S$~\cite{Nagaosa2010RMP, Ohuchi2018NatCommun}.
To quantitatively support this picture, we calculated the intrinsic anomalous Hall conductivity $\sigma_{xy}$, which is determined by the momentum-space integration of the Berry curvature. 
As shown in Figure~\ref{Fig_4}h, $\sigma_{xy}$ exhibits a pronounced variation near the relevant doping range, with a peak around $n \sim -0.04$\,e/f.u. arising from band crossings close to the Fermi level. 
This pronounced doping dependence shows that a modest polarization-induced change in the interfacial electronic structure can substantially modify the intrinsic Hall response without an appreciable change in the total magnetization. 
The calculated trend is consistent with the experimentally observed modulation of $R_\mathrm{AHE}^{r}$, supporting a mechanism in which FE polarization controls the AHE through interfacial charge redistribution and Berry-curvature modification. 
Beyond the modulation of the anomalous Hall response, the polarization-dependent reconstruction of the interfacial electronic landscape provides an additional degree of freedom for controlling the chiral spin textures and tuning their associated topological Hall responses.

Finally, the independently switchable magnetic and ferroelectric bistabilities enable four nonvolatile remanent states, offering a pathway to single-cell two-bit storage~\cite{Cao2020_NMSM}. 
Voltage ($\pm 8\,\mathrm{V}$) and magnetic-field ($\pm 0.2\,\mathrm{T}$) pulses reproducibly switch the device among four distinguishable resistance levels (Figures~\ref{Fig_4}b, i), which, although unequally spaced, exhibit robust retention over thousands of seconds (Fig.~S20 in the Supporting Information). 
The four-state response emerges below approximately 40--50\,K.
This temperature limitation does not imply a loss of ferroelectric or magnetic order, but rather a reduced efficiency in electrical readout at higher temperatures, where altered transport mechanisms diminish the required resistance contrast. 
The enhanced low-temperature resistance contrast likely stems from hopping transport, whereas altered high-temperature transport mechanisms diminish state distinguishability, though exact microscopic origins require further study. 
At still higher temperatures, however, the reduced stability of the chiral magnetic textures and ultimately the finite $T_\mathrm{c}$ of few-layer FGT impose additional intrinsic magnetic constraints.
Consequently, achieving room-temperature operation requires concurrently lifting both the transport and magnetic temperature limitations.
The rich degrees of freedom at vdW interfaces offer practical strategies to achieve this, such as modulating the FGT layer via strain, electrostatic doping, and proximity coupling, or by pairing the IS ferroelectric layer with alternative high-$T_{\mathrm{C}}$ vdW magnets. Despite current cryogenic requirements, this work successfully demonstrates the electrical tunability of emergent topological fields, highlighting FE/FM heterostructures as a highly promising paradigm for future nonvolatile multistate memory.

\section{Conclusions}

In summary, we demonstrate that FGT/IS van der Waals heterostructures host two topological Hall components associated with interfacial chiral spin textures, providing a versatile platform for their electrical control.
The thickness-dependent suppression of the dual THE signals, together with MFM imaging, atomistic spin-dynamics simulations, and first-principles calculations, provides strong evidence for chiral spin textures stabilized by interfacial DMI.
Furthermore, the ferroelectric IS layer enables nonvolatile electrical modulation of both anomalous and topological Hall responses.
By leveraging the independent bistability of ferroelectric and ferromagnetic orders, we demonstrate a nonvolatile four-state memory functionality.
These results highlight the potential of 2D FM/FE heterostructures for electrically tunable topological spintronics and multibit memory applications.

\section{Experimental Section}

\textit{Electrical Measurements:}  All electrical transport measurements were conducted using an Oxford Instruments SpectromagPT system, with a base temperature of 1.6\,K and magnetic fields up to 6\,T. A constant 500\,nA AC current at 13.333\,Hz was applied using standard low-frequency lock-in techniques. A Stanford Research Systems SR830 lock-in amplifier was used to supply the AC current and measure voltage signals. A low-noise SR560 voltage preamplifier was employed to reduce voltage noise. Gate voltages were applied using a Keithley 2614B SourceMeter.

\textit{RMCD Measurements:} RMCD measurements were performed in an attoDRY2100 closed-cycle cryostat with a base temperature of 1.5\,K and magnetic fields up to 9\,T. A power-stabilized 532\,nm Nd:YAG laser with an output power of approximately 30\,$\upmu$W was focused to a $\sim$1\,$\upmu$m beam spot on the sample at normal incidence. An optical chopper and a photoelastic modulator were used to modulate the laser intensity at frequencies $f_\mathrm{DC}$ and $f_\mathrm{PEM}$, respectively. The reflected signal was collected using a preamplified photodetector and demodulated by two synchronized lock-in amplifiers.

\textit{MFM Measurements:} MFM measurements were performed using a commercial microscope (attoAFM, attocube) installed in a closed-cycle dilution refrigerator system (Bluefors) equipped with a 9-3-1\,T vector magnet. The scanning probe system was operated at the resonance frequency of the magnetic tip, approximately 85\,kHz. The MFM images were taken in constant height mode with the scanning plane nominally $\sim$200\,nm above the sample surface. The MFM signal, i.e., the resonance frequency shift ($\Delta f$), is proportional to the out-of-plane stray field gradient. Red (blue) magnetic contrast indicates an attractive (repulsive) interaction between the magnetic tip and the sample.

\textit{Data Processing:} Due to the use of pre-patterned contacts, it is challenging to control the geometry of our devices, resulting in a mixture of longitudinal ($R_\mathrm{xx}$) and transverse (Hall, $R_\mathrm{xy}$) resistance components in both longitudinal and Hall measurements. To accurately extract these components, we applied magnetic field and geometric symmetrization techniques to isolate the individual elements of the resistance tensor. This approach relies on the principle that the longitudinal resistance is symmetric under time reversal, while the Hall resistance is antisymmetric~\cite{Sample1987}.
Specifically, above the coercive field, resistance measurements at opposite magnetic fields can be combined as follows:

\begin{equation}
	R_\mathrm{L} = \frac{R_\mathrm{meas}(B) + R_\mathrm{meas}(-B)}{2}, 
\end{equation}
\begin{equation}
	R_\mathrm{H} = \frac{R_\mathrm{meas}(B) - R_\mathrm{meas}(-B)}{2}.
\end{equation}

Here, $R_\mathrm{L}$ represents the symmetric (longitudinal) component, while $R_\mathrm{H}$ denotes the antisymmetric (Hall) component of the measured resistance, and $R_\mathrm{meas}$ represents the resistance of our devices. In our analysis, $R_\mathrm{H}$ is treated as the effective Hall resistance.

\textit{First-Principles Calculations:} The first-principles calculations were performed with the projector augmented wave method as implemented in Vienna $ab$ initio simulation package (VASP)~\cite{KRESSE199615, PhysRevB1996Kresse}. The exchange correlation effects were treated by the local density approximation~\cite{PhysRevLett1980Ceperley, PhysRevB1981Perdew}. Kohn-Sham single-particle wave functions were expanded in the plane wave basis set with a kinetic energy cutoff at 500\,eV. The energy and force convergence criteria were $10^{-7}$\,eV and $10^{-3}$\,eV/$\si{\angstrom}$, respectively. Long-range correlation is included in evaluating van der Waals interaction by the optB86b method~\cite{PhysRevB2011Klimes}. A 15×15×1 $\Gamma$-centered k-point mesh was used for Brillouin-zone integration.

For magnetic exchange calculations, we employed the generalized Bloch theorem in the reciprocal space to consider spin spirals of any wave-vector $\textit{\textbf{q}}$ for the system. We first computed the energy dispersion of homogeneous flat spin spiral excluding SOC to fit the exchange interaction parameter $J_{ij}$, and then calculated the spin spiral energy dispersion with SOC via the qSO method~\cite{PhysRevB2020Liang} to fit the DMI parameter $D_{ij}$, which treats SOC in a self-consistent way within the frame of first-order perturbation theory. We considered a cycloidal spin spiral configuration for FGT/IS heterostructure, characterized by a rotation axis $\textit{\textbf{R}} = (0,1,0)$. The spin moment $\textit{\textbf{S}}_i$ position $\textit{\textbf{R}}_i$ is described by $\textit{\textbf{S}}_i = [\cos(\textit{\textbf{q}} \cdot \textit{\textbf{R}}_i), 0 , \sin(\textit{\textbf{q}} \cdot \textit{\textbf{R}}_i)]$, where the spiral wave vector $\textit{\textbf{q}} = (q, 0, 0)$ is orthogonal to $\textit{\textbf{R}}$.

\textit{Atomistic Spin-Dynamics Simulations:} For the atomistic spin-dynamics simulations, we solved the atomistic Landau-Lifschitz-Gilbert equation with open boundary conditions. The effective Hamiltonian is written as 

\begin{equation}
\begin{split}
H = & -\sum_{i, j} J_{ij} \bm{S}_i \cdot \bm{S}_j 
- \sum_{i, j} \bm{D}_{ij} \cdot (\bm{S}_i \times \bm{S}_j) 
- \sum_i K_i (S_i^{z})^2 \\
& + \frac{1}{2} \frac{\mu_{0}}{4\pi} \sum_{i, j} \mu_{i} \mu_{j} 
\frac{(\bm{S}_i \cdot \bm{r}_{ij})(\bm{S}_j \cdot \bm{r}_{ij}) - \bm{S}_i \cdot \bm{S}_j}{\bm{r}_{ij}^{3}} \\
& - \sum_i \mu_i \bm{B} \cdot \bm{S}_i,
\end{split}
\end{equation}

which includes (i) the exchange interaction $J_{ij}$, (ii) the Dzyaloshinskii-Moriya interaction $D_{ij}$, (iii) the single-ion magnetic anisotropy $K_i$, (iv) the dipolar interactions, where $\mathbf{r}_{ij}$ denotes the unit vector of the bond connecting two spins, and (v) the Zeeman term describing the interaction of the spins with the external magnetic field $\bm{B}$. The thermal fluctuation is introduced with a stochastic field, the temperature and the time interval were set to 2\,K and 10\,fs, respectively. To reach equilibrium, 1\,ns of relaxation was performed at each magnetic field point along the hysteresis loop.

\textit{Calculations of Anomalous Hall Conductivity:} We performed first-principles calculations using VASP to obtain the band structures with the interface to the WANNIER90 package~\cite{Pizzi_2020}. A 15×15×1 k-point mesh was used in the self-consistent calculation. We selected 42 Wannier functions, using the $d$ orbitals of Fe atoms and the $p$ orbital of Te atoms, as the projection basis to obtain maximally localized Wannier functions and compute the anomalous Hall conductivity.

\subsection*{Author Contributions}

G.Y. conceived the study; Y.H. prepared the samples with the support from R.D., D.Z., J.H. and W.X.; J.X. performed the transport measurements under the instruction of R.D., A.S.M. and G.Y. and with the support from S.J., and F.L.; J.G. and R.Z. performed the MFM measurements under the instruction of S.Z. and H.G.; J.J. performed theoretical calculations under the instruction of H.Y.; G.M. performed the RMCD measurements with the support from J.X.; K.W. and T.T. grew the BN crystals; J.W. grew the FGT crystals; J.X. and Y.H. analyzed the experimental data with the help from R.D., A.S.M., S.Z., L.W. and G.Y.; J.X. and Y.H. wrote the manuscript with input from G.Y., L.W., R.D., H.D., G.C. and S.Z., and all authors contributed to the discussions and commented on the manuscript.

\subsection*{Acknowledgments}

The authors would like to thank the International Joint Lab of 2D Materials at Nanjing University for the support, and Yuanchen Co, Ltd (http://www.monosciences.com) for the High-Quality 2D Material Transfer System. G.Y. acknowledges the financial support from the National Key R$\&$D Program of China (Nos. 2024YFB3715400, 2022YFA1204700, and 2021YFA1400400), the National Natural Science Foundation of China (No. 11974169 and 12550404), the Natural Science Foundation of Jiangsu Province (No. BK20233001), and the support from Nanjing University International Collaboration Initiative. L.W. acknowledges the National Key Projects for Research and Development of China (Nos. 2022YFA1204700 and 2021YFA1400400), Natural Science Foundation of Jiangsu Province (Nos. BK20220066 and BK20233001). R.D. acknowledges the grant from the National Natural Science Foundation of China (No. 12004173). H.G. acknowledges the National Natural Science Foundation of China (No. 62488201). S.Z. acknowledges the National Natural Science Foundation of China (No. 12374199), the Beijing Nova Program (Nos. 20240484651). J.G. acknowledges the funding from the Postdoctoral Fellowship Program of CPSF (No. GZC20252200).

\subsection*{Conflicts of Interest}

The authors declare no conflicts of interest.

\subsection*{Financial disclosure}

None reported.

\subsection*{Data Availability Statement}

The data that support the figures within this paper are available from the corresponding author upon reasonable request. Source data are provided with this paper. 

\bibliography{ISFGT}

\subsection*{Supporting Information}

Additional supporting information can be found online in the Supporting Information
section.


\end{document}